\documentclass[fleqn,twoside]{article}
\usepackage{espcrc2}

\usepackage{graphicx}
\usepackage[figuresright]{rotating}

\newcommand{\AmS}{{\protect\the\textfont2
  A\kern-.1667em\lower.5ex\hbox{M}\kern-.125emS}}

\title{Investigating QCD Vacuum on the lattice}

\author{A. Di Giacomo\address[MCSD]{Dipartimento di Fisica and I.N.F.N., 
        \\ 
        Via Buonarroti 2, 56100 Pisa, Italy}%
        \thanks{Partially supported by MURST and by EC, FMRX-CT97-0122},
        }
       
\begin{document}

\begin{abstract}
Investigations on the structure of QCD vacuum from first principles can be done
on the lattice. The mechanism of confinement is an example: results from lattice
on it are reviewed.
\vspace{1pc}
\end{abstract}

\maketitle

\section{Introduction}
Euclidean Feynman path integral uniquely identifies the ground state of a
field theory (vaccum)\cite{1}.

The Feynman functional integral is defined as the limit of ordinary
integrals defined on discrete set of points in a four dimensional box,
when the number of points is sent to infinity filling the box densely;
the size of the box is then sent to infinity to cover the whole space
time. Lattice formulation is an approssimant in this sequence. 

QCD has an
UV fixed point at $g=0$ (asymptotic freedom): as $g\to 0$ the physical
length scale $\xi$ increases in units of the lattice spacing $a$
\begin{equation}
\frac{a}{\xi} \mathop\rightarrow_{g_0\to 0} = 0
\end{equation}
The density of lattice points in physical units goes large. On the other
hand a mass gap exists in the theory, which makes the infinite volume
limit well defined as a thermodynamical limit.

Lattice is a good approximant of QCD if
\begin{equation}
a \ll \xi \ll La\label{eq2}
\end{equation}
with $L$ the size of the lattice.

The above argument is a strong indication that QCD most probably exists as
a field theory in the constructive sense, and can be defined as a limit of
lattice formulation as $g_0\to 0$, and $L$ is such that the inequality
(\ref{eq2}) is satisfied.

The textbook quantization of QCD is based on perturbation theory, and the
ground state is Fock vacuum of quarks and gluons.

Phenomenological evidence exists\cite{2} that the QCD vacuum is not the
perturbative vacuum. The instability of Fock's vacuum is possibly the
origin of the non Borel summability of the perturbative expansion\cite{2a}.

Perturbation theory apparently works at small distances but is not well
defined and is unable to describe large distance physics.

The most clear evidence that Fock vacuum is not a good approximation to
QCD ground state, is that its elementary excitations, quarks and gluons,
have never been observed as free particles. This phenomenon is known as
confinement of colour and is one of the most intriguing properties of QCD.

The study of the mechanism of confinement is an important chapter of the
 investigation of the structure of QCD vacuum on the lattice, and will be
the object of this talk. In sect.2 I will review the experimental evidence
for confinement. In sect.3 I will discuss the phenomenology of the
deconfinement transition as observed in numerical simulations on the
lattice. This will naturally lead us to the idea of duality, which will
be the object of sect.4. Progress in understanding confinement as dual
superconductivity of the vacuum will be  reviewed in sect.5
\section{Confinement: experimental evidence.}
Confinement is defined as the absence of colored particles in asymptotic
states.

The existing experimental evidence for confinement is based on the
negative result of searches of fractionally charged particles (quarks) in
particle reactions and in nature.

The cross section $\sigma_q$ for the inclusive production of quark or
antiquark in the process
\[ p + p \to q (\bar q) + X\]
at c.m. energies $\sim 100$~GeV  has an upper limit\cite{3}
\[\sigma_q < 10^{-40}\,\rm{cm}^2\]
to be compared with the total cross section $\sigma_T$ at the same energy
$\sigma_T \sim 10^{-25}\,{\rm cm}^2$.

The ratio $R_q = \sigma_{q}/\sigma_{T}$ has then the upper bound
\[ R < 10^{-15}\]
while the expected value in the absence of confinement is a sizable
fraction of unity.

The negative result of the search of fractionally charged particles in
ordinary matter by Millikan-like experiments gives an upper limit for the
ratio $R$ of the quarks abundance $n_q$ to nucleons abundance
$n_p$
\[ R_{ob} = \frac{n_q}{n_p} < 10^{-27}\]
corresponding to the analysis of $\sim1 $~gr of matter.

In the absence of confinement the expectation for $R$, $R_{SCM}$, in the
standard cosmological model is $R_{SCM} \sim 10^{-12}$\cite{4}. Again
\[R/R_{SCM} < 10^{-15}\]
A ratio smaller than $10^{-15}$, if different from zero, 
would be too small to
have a natural explanation in any theory. The most natural interpretation
is then that those ratios are strictly zero, or that confinement is
an absolute property of the vacuum based on a symmetry\cite{5}.

The question is: what symmetry of QCD vacuum prevents quarks to exist as
free particles?

As for gluons, they have no such characteristic signature as a fractional
charge, and their identification is not clearly feasible. No experimental
data exist on gluon confinement. We shall define anyhow confinement
as absence of any colored particle as a free particle.
\section{Deconfinement Phase transition on the lattice.}
QCD at finite temperature can be studied on the lattice. The partition
function $Z(T)$ of  a field with action $S[\Phi]$ is equal to the Feynman
integral in Euclidean space, with the time direction extending from 0 to
$1/T$ and periodic boundary conditions in time for bosons, antiperiodic
for fermions
\begin{equation}
Z(T) = \int \Pi[{\cal D}\Phi] e^{
-\beta\!\int \! d^3x\int_0^{1/T}\hskip-5pt dx_0 S_E[\Phi(\vec x,x_0)]
}
\end{equation}
In lattice QCD this corresponds to having a lattice of size $N_S^3\times
N_T$, with $N_S\gg N_T$, and the temperature is given by the inverse of
the temporal extension $N_T a$ ($a$ the lattice spacing).

The value of $a$ in physical units depends on the coupling constant
($\beta = 2 N/g^2$) via renormalization group
\[
a(\beta) = \frac{1}{\Lambda_L} \exp (- b_0 \beta)\]
with
\[ b_0 > 0\qquad \hbox{(asymptotic freedom)}\]
Hence
\begin{equation}
T = \frac{1}{N_T a} = \frac{\Lambda_L}{N_T}\exp(b_0 \beta)
\end{equation}
As a consequence of asymptotic freedom low temperature (confinement)
corresponds to strong coupling (large $g$) or to disorder in the language
of statistical mechanics, high temperature corresponds to order or to weak
couling.

If confinement is due to a symmetry, it has to be a symmetry of the
disordered phase. This naturally leads to the idea of duality, which will
be the object of sect.3.

The deconfinement transition is detected on the lattice in pure
gauge theory, by looking at the correlator
\begin{equation}
{\cal D}(\vec x-\vec y) = \langle L(\vec x) L(\vec y)\rangle
\end{equation}
where $L(\vec x)$,
the Polyakov line, is the trace of the
parallel transport along the time axis across
the lattice and back via periodic boundary conditions
\begin{equation}
L(\vec x) = 
Tr\left\{P\exp\left( i \int_0^{a N_T}\hskip-3pt A_0(\vec x,x_0)
dx^0\right) \right\}
\end{equation}
The static potential between a $q\,\bar q$ pair is given by
\begin{equation}
V_{q\bar q}(\vec x - \vec y) = \frac{\ln{\cal D}(\vec x -
\vec y)}{a N_T}
\end{equation}
By cluster property, at large distances
\begin{equation}
{\cal D}(\vec x - \vec y)\simeq C\exp(-A |\vec x - \vec y| a N_T)
+ K |\langle L\rangle|^2
\end{equation}
A critical temperature $T_c$ is observed such that
\[\begin{array}{lccl}
\rm{for}\, T < T_c & \langle L \rangle = 0 & V(r)
\mathop{=}\limits_{r\to\infty}
\sigma r & \rm{(conf.)}\\
\rm{for}\, T > T_c & \langle L \rangle \neq 0 & V(r)
\mathop{=}\limits_{r\to\infty}
0 & \rm{(deconf.)}
\end{array}
\]
Confinement is related to the presence of a linear potential at large
distances: $\sigma$ is known as string tension.

For $SU(2)$ gauge theory the phase transition at $T_c$ is second
order,and belongs to the universality class of the $3d$ Ising model.
$T_c\simeq 180\,{\rm{MeV}}$\cite{6}.

For $SU(3)$ gauge theory the transition is weak first order, and $T_c
\simeq 270\,{\rm{MeV}}$\cite{7}.

In the presence of quarks the symmetry $Z_N$ of which $L$ is an order
parameter, is not a symmetry any more. For massless quarks a chiral
symmetry exists above some temperature $T_c$, which is spontaneously
broken for $T< T_c$, the pseudoscalar octet being the Goldstone
particles. For $m_q\neq 0$ the chiral symmetry is again explicitely
broken.

It is qualitatively clear that confinement can produce a breaking of
chiral symmetry: in a bag model chirality is inverted in the reflection
on the confining wall.
Numerical indications also exist that the two transitions take place at
the same temperature.
However the overall situation is not satisfactory.

First of all if a phase exist in which color is confined, and a phase
at higher temperature in which quarks and gluons are free particles, an
exact order parameter should exist for this transition.

In addition a strong theoretical hint exists that, when the number of
colors $N_c$ is sent large, with the constraint $N_c g^2 = \lambda$
fixed, a limiting theory is defined, which does not differ much in its
physical content from the realistic theory where $N_c = 3$.

The expansion in $1/N_c$ should be a convergent expansion\cite{8}.

In this philosophy quark loops are non leading ${\cal O}(1/N_c)$ and
therefore the physics of
confinement, i.e. the symmetry, should be the same as in quenched
approximation.
\section{Duality\cite{9}.}
Duality is a deep concept in field theory, in statistical mechanics, 
in string theory. It applies to
systems in $d+1$ dimensions having non local topological excitations in 
$d$ dimensions. Two complementary descriptions can be given of such systems.

A {\em direct} description, in terms of local fields $\Phi$, in which
topological excitations $\mu$ are non local. Symmetry is described by
order parameters, the v.e.v. $\langle\Phi\rangle$ of the fields $\Phi$.
This description is convenient in the ordered phase, or weak coupling
regime.

A {\em dual} description in which $\mu$ are local operators, and $\Phi$ non
local excitations. Symmetry is described by disorder parameters
$\langle\mu\rangle$. The dual coupling is $g_D\sim 1/g$. This description
maps the strong coupling regime of the direct theory into the weak coupling
regime of the dual. Therefore it is convenient in the strong coupling
regime.

The prototype system\cite{10} with duality is the $1+1$ dimensional Ising
model: there the field $\Phi$ is the spin variable $\sigma(i) = \pm1$.
The dual configurations $\mu$ are 1 dimensional kinks. The dual
description is again an Ising model in which the creation operator of a
kink is $\mu(i) = \pm 1$ and the dual Boltzman factor $\beta_D$ is
defined by the relation
\begin{equation} \sinh 2 \beta = \frac{1}{\sinh2\beta_D}
\label{eqsinh}
\end{equation}
or $\beta\sim 1/\beta_D$. In the model $T= 1/\beta$ plays the role of the
coupling constant.

Other examples are
\begin{itemize}
\item[1)] The 3d $XY$ model\cite{11}, which belongs to the class of
universality of liquid $He_4$. The topological excitations are abelian
vortices.
\item[2)] The Heisenberg magnet, where the topological excitations are
$2d$ Weiss-domains\cite{12}.
\item[3)] the $N=2$ SUSY QCD\cite{13} where the excitations are monopoles.
\item[4)] Superstring $M$ theories\cite{14}.
\item[5)] Compact $U(1)$ gauge theories, where the topological excitations
are monopoles\cite{15,16}.
\end{itemize}

The question is: what are the dual excitations in QCD? 

Two proposals exist
in the literature, both due to G. t'Hooft\cite{5,17}.
\begin{itemize}
\item[a)] {\em Monopoles}. Monopole condensation in the vacuum produces
dual superconductivity, and confinement of electric charges via Abrikosov
flux lines (Meissner effect). Monopoles are defined by a procedure named
``abelian projection''\cite{17} based on the choice of  a local operator in
the adjoint representation, as discussed below. In a sense the mechanism
is largely undefined, since there is a continuous infinity of choices for
the operator used to define the monopoles and their interrelation is not
understood. A guess is that all these choices are physically
equivalent\cite{17}.

The positive
feature is that confinement is explained in a simple form: the
chromoelectric field of a
$q\bar q$ pair is channeled into an Abrikosov flux tube with energy
proportional to the length, i.e. to the
distance $r$ between the pair\cite{18,19}
\[ V(R) = \sigma R\]
$\sigma$, the energy per unit length of the tube, is the string
tension.
\item[2)] {\em Vortices}. A vortex is a magnetic defect associated to a
closed line $C$. The operator $B(C)$ which creates a vortex at some time
$t$, obeys the following algebra, with the operator $W(C')$ creating a
Wilson loop along
the line $C'$
\[ B(C) W(C') = W(C') B(C) \exp\left(i \frac{n_{CC'}}{N}2\pi\right)\]
where $n_{CC'}$ is the winding number of the lines $C,C'$.
It can be shown that, whenever $\langle W(C')\rangle$ obeys the area law,
$\langle W(C')\rangle \sim \exp(-Area_{C'})$, as the loop goes large,
$\langle B(C)\rangle$ obeys the perimeter law, and viceversa if 
$\langle B(C)\rangle$ obeys the area law, $\langle W(C')\rangle$ obeys the
perimeter law.

Area law implies that $\langle L\rangle$, the expectation value of the
Polyakov line vanishes. One can define a dual Polyakov line $L'$ wich is a
vortex along a straight path crossing space from $-\infty$ to $+\infty$.
Then for $T< T_c$, $\langle L\rangle = 0$, for 
$T > T_c$, $\langle L'\rangle = 0$. $\langle L'\rangle$ is a disorder
parameter for confinement\cite{20}.
\end{itemize}
The definition of monopoles is associated to the choice of an operator
$\vec\Phi(x)$ in the adjoint representation: we will speak of $SU(2)$ gauge
group for simplicity, but the procedure is easily generalized to $SU(N)$.
Let $\hat\Phi(x) \equiv \vec\Phi(x)/|\vec\Phi(x)|$ be the direction of
$\vec \Phi$ in color space. A gauge invariant field strength $F_{\mu\nu}$
can be defined as\cite{20bis}
\begin{equation} F_{\mu\nu} = \hat\Phi\cdot\vec G_{\mu\nu} - \frac{1}{g}
\hat\Phi(D_\mu\hat\Phi\wedge D_\nu\hat\Phi)\label{eq11}
\end{equation}
\[ \vec G_{\mu\nu} = \partial_\mu \vec A_\nu -\partial_\nu\vec A_\mu
+ g\vec A_\mu\wedge\vec A_\nu\]
\[
D_\mu = \partial_\mu - g \vec A_\mu\wedge\]
The two terms in the definition (\ref{eq11}) are separately gauge
invariant and color singlets: they are arranged in such a way that
bilinear terms
$A_\mu A_\nu$ or $A_\mu\partial_\nu\Phi$ cancel. Indeed
\begin{equation} F_{\mu\nu} = \hat\Phi(
\partial_\mu \vec A_\nu -\partial_\nu\vec A_\mu) - \frac{1}{g}
\hat\Phi(\partial_\mu\hat\Phi\wedge \partial_\nu\hat\Phi)
\end{equation}
The gauge transformation bringing $\Phi$ in the same direction (say (0,0,1))
in color space is called abelian projection. After abelian projection
\begin{equation}
 F_{\mu\nu} = \partial_\mu  A^3_\nu -\partial_\nu A^3_\mu
\end{equation}
The source of the dual tensor $F^*_{\mu\nu} =
1/2\varepsilon_{\mu\nu\rho\sigma}F^{\rho\sigma}$ is a magnetic current
\begin{equation} j_\nu = \partial^\mu F^*_{\mu\nu}\end{equation}
and is conserved. The corresponding $U(1)$ magnetic symmetry can either be
Wigner, and then a magnetic charge operator is defined and Hilbert space is
superselected, or be Higgs broken, which implies the existence of at least
one magnetically charged operator $\mu$ such that $\langle\mu\rangle\neq 0$.
$\langle\mu\rangle\neq 0$ implies dual superconductivity. Notice that in a
noncompact formulation $j^\nu = 0$ (Bianchi identities).

In principle the existence of dual superconductivity can be investigated
in any abelian projection by looking at the v.e.v. $\langle\mu\rangle$ of an
operator creating a magnetic charge in that abelian projection.
\section{Construction of the disorder parameter
$\langle\mu\rangle$\cite{21,22,23,24}} Once the dual topological
excitations are identified the disorder parameter can be constructed as
the v.e.v. of their creation operator $\mu$. The construction of $\mu[\Phi]$
interms of the field of the direct description
$\Phi$ is an explicit realization of the duality transformation.

The guiding idea goes back to ref.\cite{25} and amounts to a traslation of
the fields in the Schr\"odinger picture by the classical topological
configuration. In the same way as
\begin{equation} e^{ipa} | x\rangle = |x + a\rangle
\label{eqexp}\end{equation}
\begin{equation}\mu(x) = \exp\left(
i\int d^3y \Pi(x_0,\vec y)\bar\varphi(\vec y-\vec x)\right),
\label{eqmu}\end{equation}
with $\Pi$ the conjugate momentum to the field $\Phi$
\[\left[ \Phi(x_0,\vec x), \Pi(x_0,\vec y)\right] =
i \delta^3(\vec x - \vec y),
\]
adds $\bar\varphi(\vec x)$ to the field configuration
\[
\mu(\vec x)|\Phi(\vec x)\rangle =
|\Phi(\vec x) + \bar\varphi(\vec x)\rangle\]
Adapting the above construction to a compact formulation of the theory as
in QCD on the lattice is far from trivial, but has been
done\cite{22,23,24}.

A correct,gauge invariant, definition of the operator $\mu$ for
abelian projected monopoles exists\cite{22}. The resulting operator is a
Dirac like magnetically charged and gauge invariant operator, which can
then have non vanishing v.e.v. without violating gauge invariance\cite{26}.

The construction has been checked on systems already studied and
understood by other methods, starting with the 2d Ising model\cite{27},
where our operator
$\mu$ which creates kinks does indeed coincide with the dual variable
$\sigma^*$ of ref.\cite{10}.

For the 3d $XY$ model the phase transition was always conjectured to be
produced by condensation of vortices, but that condensation was detected
in numerical simulations by observing an increase of the number of
vortices in condensed phase: our construction shows that
$\langle\mu\rangle\neq0$, or that the number of vortices is not defined
in the vacuum of the condensed phase\cite{11}.

For the Heisenberg model we have found\cite{12} that the Curie transition
can be looked as an order disorder transition, and that the
high temperature phase is ordered in the dual language, by condensation of
non abelian
$O(3)$ vortices. 

A numerical problem has also been solved; $\mu$ defined by eq.(\ref{eqmu})
is the exponential of an integral over the spatial volume: therefore it
fluctuates typically as $\exp(L^{3/2})$, which is a wild fluctuation.
In fact the disorder parameter $\langle\mu\rangle$ is the ratio of two
partition functions
\[ \langle\mu\rangle = \frac{\tilde Z}{Z}\]
and has the fluctuations typical of a partition function. $Z$ is the
ordinary partition function of the theory, and $\tilde Z = \int e^{-\tilde
S}$ corrisponds to a modification of the action coming from the exponent in
$\mu$, eq.(\ref{eqmu}).

A way out of this difficulty consists\cite{28,21} in considering instead
of
$\langle\mu\rangle$ the quantity
\begin{equation}
\rho = \frac{d}{d\beta} \ln \langle\mu\rangle =
\langle S\rangle_S - \langle {\tilde S}\rangle_{\tilde S}
\end{equation}
$\beta = \frac{2N}{g^2}$,
which contains all the relevant information of $\langle\mu\rangle$.

A typical shape of $\langle\mu\rangle$ is shown in fig.1. The corresponding
$\rho$ is shown in fig.2. The peak seats  on the location of the phase
transition.

\begin{figure}[htb]
\includegraphics[width=0.9\linewidth]{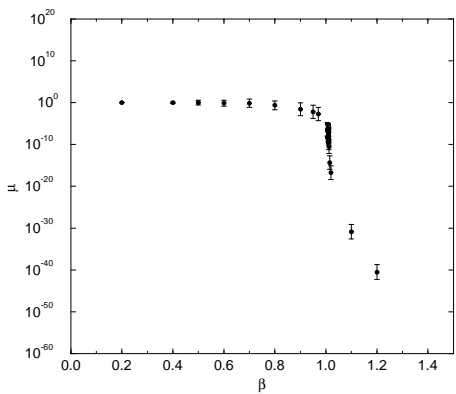}
\caption{$\langle\mu\rangle$ for compact $U(1)$.}
\end{figure}

\begin{figure}[htb]
\includegraphics[width=0.9\linewidth]{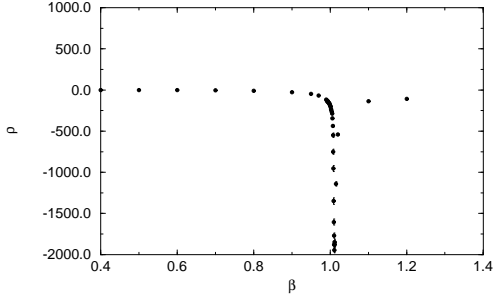}
\caption{$\rho$ for compact $U(1)$}
\end{figure}

The phase transition in principle takes place at infinite volume. The limit can
be performed by a finite size scaling analysis as follows; if $\delta$ is the
critical index of $\mu$ and $\tau = (1- T/T_c)$ the reduced temperature,
\[ \langle\mu\rangle \mathop\simeq_{\tau \to 0}
\tau^\delta\]
then
\begin{equation}
\langle\mu\rangle \mathop\simeq_{\tau \to 0}
\tau^\delta\Phi(\frac{a}{\xi}, \frac{L}{\xi})
\end{equation}
The functional dipendence is dictated by the fact that $\Phi$ is
dimensionless: $a$ is the lattice spacing, $\xi$ the correlation
length, $L$ the extension of the lattice. As $\tau\to 0$, $\xi$ goes
large with a critical index $\nu$
\[ \xi \propto \tau^{-\nu}\]
 $a/\xi$ can be approximated with 0 (scaling limit), and $L/\xi$
can be traded with $L^{1/\nu}\tau$. Hence
\[ \langle\mu\rangle \sim \tau^\delta f(L^{1/\nu}\tau)\]
and
\begin{equation} \rho L^{-1/\nu} = \frac{\delta}{\tau L^{1/\nu}} +
f'(L^{1/\nu}\tau)
\label{eqrho2}\end{equation}
$\rho/ L^{1/\nu}$ is  a universal function of $L^{1/\nu}\tau$, i.e. it
scales when plotted versus $L^{1/\nu}\tau$.

Since this scaling law is valid for the appropriate value of $\nu$ and
$\beta_c$, the determination of $\nu$, $\beta_c$ and $\delta$ is then possible.
\section{Results for QCD\cite{22,23,24}.}
$\langle\mu\rangle$, or better $\rho$, has been studied for pure gauge 
$SU(2)$ and $SU(3)$ gauge theories across the deconfining phase transition,
in a number of different abelian projections.

A clear evidence has been obtained that, irrespective of the choice of the
abelian projection
\begin{itemize}
\item[1)] $\langle\mu\rangle\neq 0$ in the confined phase ($T < T_c$).
\item[2)] $\langle\mu\rangle \sim \exp(-k L_S)$, $k > 0$ for $T> T_c$,
$L_S$ being the spatial extension of the lattice. This means
$\langle\mu\rangle = 0$ at $T > T_c$ in the thermodynamical limit.
\item[3)] 
\[ \langle\mu\rangle \mathop\sim_{T\to T_c^-} 
\left(1 - \frac{T}{T_c}\right)^\delta\]
\item[4)] $T_c$ and $\nu$ can be determined, together with $\delta$ from the
finite size scaling analysis sketched in sect.3.
\end{itemize}
Fig.3 shows a typical form of $\rho$ as a function of $\beta$ for different
lattice sizes and $SU(2)$ gauge theory.

Fig.4 shows the quality of the scaling equation (\ref{eqrho2}), i.e. the
validity of the extrapolation at infinite volume.

\begin{figure}[htb]
\includegraphics[width=0.9\linewidth, angle=270]{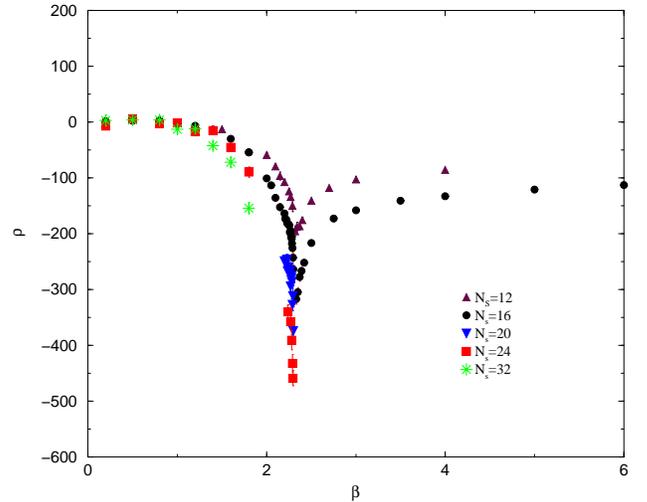}
\caption{$\rho$ for monopole condensation in $SU(3)$.}
\end{figure}

\begin{figure}[htb]
\includegraphics[width=0.9\linewidth]{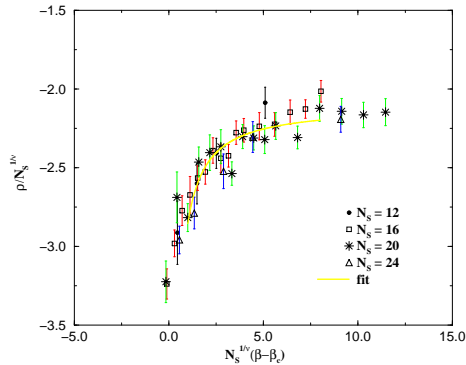}
\caption{Finite size scaling of $\rho$.}
\end{figure}

$T_c$ agrees with the determination made using the traditional order
parameter $\langle L\rangle$, and so does $\nu$. For $SU(2)$ $\nu =
0.62(2)$ in agreement with the expectation that the transition belongs to
the universality class of the 3d Ising model, $\delta = .20(8)$.

For $SU(3)$, $\nu = 1/3$, which means that  the transition is first order,
even if weak enough to allow a scaling region, and $\delta = .50(3)$.

The independence of the result from the choice of the abelian projection
has been checked by performing the analysis in a number of different
abelian projections, but also averaging on a very large number 
(infinite)
of abelian projections\cite{24}.

These results clearly indicate that confining vacuum is  a dual
superconductor.

The same disorder parameter $\langle\mu\rangle$ can be used in full $QCD$
(including dynamical quarks) where it is well defined, contrary to $\langle
L\rangle$ whose symmetry $Z_N$ is explicitely broken by the very presence
of quarks and to $\langle\bar\psi\psi\rangle$, whose symmetry is broken by
quark masses.

The analysis for full QCD is in progress and of course requires a big
computational effort.
There the dependence on the masses of the quarks makes the finite size
scaling analysis, eq.(\ref{eqrho2}) more complicated. Preliminary
results\cite{29}, however, indicate that also in full QCD there is a transition
from dual superconductor to normal across the transition. If these preliminary
indications get confirmed by the quantitative analysis which is on the way,
we would have a good order parameter for confinement, even in  the presence of
(massive) quarks.
In principle the two transitions, chiral and 
deconfinement, could take place at different temperatures; 
however all the existing indications are that they coincide.

Moreover this
would also
reconcile confinement with the $N_c\to \infty$ limit. In that respect also an
analysis of quenched gauge theories at $N_c > 3$ is on the way, to perform the
brute force check of the limit.

Here again preliminary results show that the order parameter is weakly
dependent on $N_C$ at $g^2 N_C = const.$\cite{30}.

However our results are still far from complete. They indicate that,
whatever the dual fundamental excitations are, they carry magnetic charge
in all the abelian projections. A real theoretical breakthrough would be to
identify such excitations, and to write an effective Lagrangian for them to
describe QCD in the confined phase. 

Our results will hopefully help in solving
this problem.

I thank L. Del Debbio, M. D'Elia, B. Lucini, G. Paffuti, for discussions.
Most of the work reported here is largely due to their collaboration.


\begin{thebibliography}{99}
\bibitem{1} R.P. Feynman, Rev. Mod. Phys. 20 (1948) 367.
\bibitem{2} M.A. Shifman, A.I. Veinshtein, V.I. Zakharov, Nucl. Phys.
B147 (1979) 385,448,519.
\bibitem{2a} A.H. M\"uller, Nucl. Phys. B250 (1985) 327.
\bibitem{3} Review of Particle Physics, E.P.J. 15 (2000).
\bibitem{4} L. Okun, Leptons and quarks, Norh Holland (1982).
\bibitem{5} G. t'Hooft, Nucl. Phys. B138 (1978) 1.
\bibitem{6} J. Engels, F. Karsch, H. Satz, I. Montway, Nucl. Phys. B205
(1982) 239.
\bibitem{7} B. Beinlick, F. Karsch, E. Laerman, A. Peikart, Eur. Phys. J.
C6 (1999) 133.
\bibitem{8} G. t'Hooft, Nucl. Phys. B72 (1974) 461.
\bibitem{9} H.V. Kramers, G.H. Wannier, Phys. Rev. 60 (1941) 252.
\bibitem{10} L. Kadanoff, H. ceva, Phys. Rev. B3 (1971) 3918.
\bibitem{11} G. Di Cecio, A. Di Giacomo, G.Paffuti, M. Trigiante, Nucl.
Phys. B489 (1997) 739.
\bibitem{12} A. Di Giacomo, D. Martelli, G. Paffuti, Phys. Rev. D 
(1999) 094511.
\bibitem{13} N. Seiberg, E. Witten, Nucl. Phys. B431 (1994) 484.
\bibitem{14} 
T. Banks, W. Fischler, S.H. Shenker, L. Susskind:
{ Phys. Rev.} { D55}, (1997), 5112.
\bibitem{15} J. Fr\"ohlich, P.A. Marchetti, Comm. Math. Phys. 112 (1987)
343.
\bibitem{16} A. Di Giacomo, G. Paffuti, Phys. Rev. {  D56} (1997) 6816 .
\bibitem{17} G. t'Hooft, Nucl. Phys. B190 (1981) 455.
\bibitem{18} G. t'Hooft, High Energy Physics, EPS International
Conference,
Palermo (1975), eds A. Zichichi.
\bibitem{19} S. Mandelstam, Phys. Rev. 23C (1976) 245.
\bibitem{20} L. Del Debbio, A. Di Giacomo, B. Lucini, Nucl. Phys. B594
(2001) 287.
\bibitem{20bis} G. t'Hooft, Nucl. Phys. B79 (1974) 276.
\bibitem{21} L. Del Debbio, A. Di Giacomo, G. Paffuti, P. Pieri, Phys.
Lett. B355 (1995) 255.
\bibitem{22} A. Di Giacomo,, B. Lucini, L. Montesi, G. Paffuti,
Phys. Rev. D61 (2000) 034500.
\bibitem{23} A. Di Giacomo,, B. Lucini, L. Montesi, G. Paffuti,
Phys. Rev. D61 (2000) 034505.
\bibitem{24} J. M. Carmona, M. D'Elia, A. Di Giacomo,, B. Lucini, G.
Paffuti, Phys. Rev. D64 (2001) 114507.
\bibitem{25} E. Marino, B. Schroer, J. A. Swieca, Nucl. Phys. B200 (1982)
473.
\bibitem{26} A. Di Giacomo, G.Paffuti, 19th International Symposium on
Lattice Field Theory, Berlin 2001, hep-lat 0110061, to appear in the
Proceedings.
\bibitem{27}J. M. Carmona,  A. Di Giacomo,, B. Lucini, Phys. Lett. B485
(2000) 126.
\bibitem{28} L. Del Debbio, A. Di Giacomo, G. Paffuti, Phys. Lett. B349 (1995)
513.
\bibitem{29} J. Carmona, M. D'Elia, L. Del Debbio, A. Di Giacomo, B. Lucini,
G. Paffuti, contribution to Lattice 01 Berlin, hep-lat 010058.
\bibitem{30} L. Del Debbio, A. Di Giacomo, S. Betti, in preparation
\end{thebibliography}
\end{document}